# CLINICAL AND EDUCATIONAL APPLICATIONS OF LIVEIA: AN IMMERSIVE VISUALIZATION ENVIRONMENT

**Liane Gabora**
*Department of Psychology, University of British Columbia*

**Abstract**

The association between light and psychological states has a long history and permeates our language. LIVEIA (Light-based Immersive Visualization Environment for Imaginative Actualization) is a new immersive, interactive technology that uses physical light as a metaphor for visualizing peoples' inner lives and relationships. This paper outlines its educational value, as a tool for understanding and explaining aspects of how people think and interact, and its potential therapeutic value as a form of art therapy in which the artwork has straightforwardly interpretable symbolic meanings.

## 1. Introduction

The association between light and psychological states dates back to the 'dawn' of civilization (Zajonc, 1993). It appears in numerous religious traditions and permeates our language; for example, creative *spark*, moment of *illumination*, *flash* of insight, *brilliant* idea, *bright* versus *dim*-witted person, a ray of hope, and enlightened state. LIVEIA (Light-based Immersive Visualization Environment for Imaginative Actualization') is a new interactive technology that uses physical light as a metaphor for visualizing peoples' inner lives (Gabora, 2014, 2015). Visualization has proven effective for facilitating understanding of everything from weather patterns to stock market trends, but its potential to facilitate understanding of psychological phenomena is virtually untapped. LIVEIA uses properties of light and principles of optics (Holtmannspötter, & Reuscher, 2009; Reinhard, Khan, Akyuz, & Johnson, 2008; Valberg, 2005) to enable people to systematically depict the elusive aspects of human nature and mental processes, as well as specific situations they may find themselves in, using symbolic representations of themselves and others that can be explored and experimented with. This short paper will discuss its potential therapeutic value as a form of art therapy in which the artwork has straightforwardly interpretable symbolic meanings, and its educational value, as a tool for understanding and explaining many aspects of how people think and interact, including patterns of communication and miscommunication, how they come about, and how they could unfold over time.

## 2. Design

*Light Metaphor Underlying the Technology*. In LIVEIA, physical light is understood to represent 'inner light', which can refer to creative spark, life force, spiritual essence, or chi. An individual's psyche is represented by a sphere that amplifies and transforms light, as shown in **Fig. 1a**. Users are shown how to use spheres to represent themselves and people with whom they interact, and instructed how to translate peoples' attributes—such as personality traits, areas of expertise, and so forth—into visible attributes of the spheres. For example, vibrant people are portrayed using a high intensity sphere while an individual with little life force is represented as having low light intensity, as shown in **Fig. 1b**. Accessible, transparent people are depicted with transparent, thin-shelled spheres, while aloof people, or those who hide their true nature, depicted with opaque spheres that locally trap their light as shown in **Fig. 1c**.

*Figure 1. (a) A sphere of a material with a higher refractive index than air (e.g., crystal) traps and amplifies light (left). A sphere may appear dark because (b) its light is of low intensity (center), or (c) its light is obscured (right).*

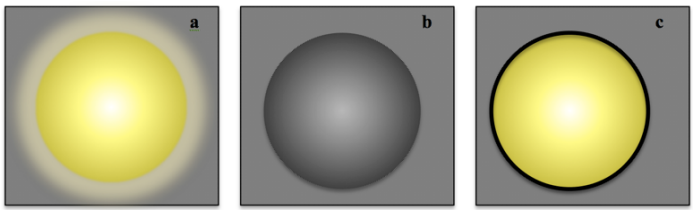

The more comfortable people feel with each other the more blurred their sphere surfaces when in close proximity. Thoughts are represented as beams of light with particular waveforms. Complex thoughts are depicted as superpositions of simpler waveforms. The greater the emotional valence associated with a thought, the more intense the beam. The vaguer or less well understood a thought or idea, the more diffuse the beam. Superficial personality traits are represented by color patterns on the sphere surface.

## 3. Therapeutic and Educational Applications

LIVEIA can be used as an educational tool for explaining aspects of how people think and interact, or as a form of art therapy in which the depiction has straightforwardly interpretable symbolic meanings. Perception of external stimuli is represented using beams of light that are intercepted by the sphere. Thoughts and ideas are represented as beams of light that originate within the sphere. The experience of being unable to clearly articulate what one is thinking or feeling is represented using a diffuse beam of light that diverges as it passes out of the sphere, as illustrated in **Fig. 2a**. The less focused it is, the more it diverges.

*Figure 2. (a) Because the inner surface of a sphere is concave, a beam of light diverges (becomes less focused) when it passes through the sphere. (b) When a beam (lower left) reaches the interior surface of a sphere, it breaks into two: a reflected beam (which returns into the sphere), and a refracted beam (which bends as it passes through). Due to the concavity of the sphere, the refracted ray diverges as it passes through, while the reflected ray converges (becomes more focused). (c) A beam (upper left) is distorted (scatters, reflects, refracts, or all three) when it encounters a fracture. A fractured sphere may contain shadowy regions that light cannot penetrate. (d) Light that originates from the center can radiate in any direction without refraction because wherever it contacts the surface it is perpendicular to it.*

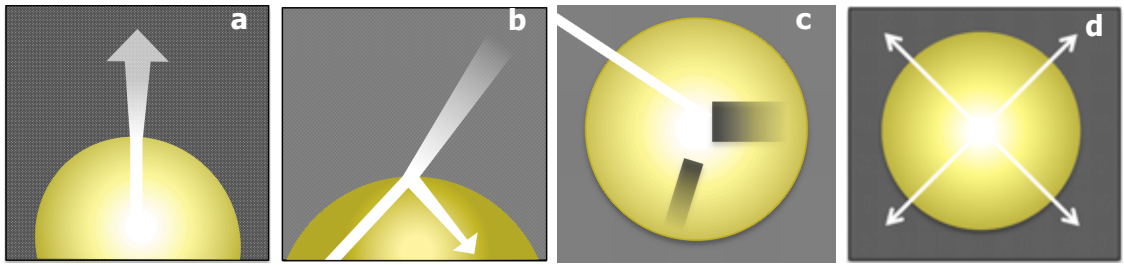

*Reflection on an Experience or Idea*. It is also possible to depict how one can find a way of expressing a particular thought or emotion through the process of considering it from different perspectives, or *reflecting on* it. Reflection is literally depicted as reflection of a beam of light, as shown in **Fig. 2b**. When reflected off the concave inner surface of the sphere, diffuse beams—representing vague thoughts and ideas—converge and come into focus. The user can thus document the process by which a particular problem is "bounced around in the mind" until it becomes sufficiently focused to be articulated and understood. The refracted ray depicts the perceivable signs (such as a facial expression of concentration) that one is engaged in focused reflection on a problem. Vague ideas are represented as diffuse beams that require much reflection. A thought or experience that is no longer being reflected upon is depicted as a spark of light within the sphere.

*Deception*. The proclivity to deceive others is highly correlated with a distorted perception of reality (Beck, 1990), and this too can be depicted in LIVEIA. A fracture (or impurity or vein of different material) will cause a beam of light to bend (refract), and change direction. Thus deception, i.e., bending the truth, is represented as the deliberate redirection of a beam of light through a fracture. In other words, bending the truth is represented as the bending of a beam of light. Lying to someone else is depicted as occuring at surface of the sphere, while lying to oneself is depicted as occurring within the sphere. Aspects of oneself or one's life that one wants to avoid, i.e., the "shadow side" of the psyche, are depicted as fractured regions of the sphere that obstruct the natural flow of light, as illustrated in **Fig. 2c**. Fracturing can represent, literally, a lack of integrity, a state wherein one is living with lies, or where one's values are not in sync with ones' actions, or one is living with memories that are too painful to face.

*Enlightenment*. A state of enlightenment is depicted as a sphere without fractures, such that the interior is uniformly lit, as in **Fig. 2d**. Since there are no fractures blocking light from the core, an enlightened individual can communicate from the core of the self, as opposed to a superficial level. Mindfulness is depicted as the state of remaining alert to the presence of fragmentation or shadows and reflecting on them—considering them from different perspectives—to overcome them.

Users can depict scenarios involving one or more individuals, re-orient scenarios to get a new perspective on them, and make copies of scenarios to explore how they arose or what might happen next.

## 4. Discussion and Conclusions

This paper has skimmed the surface of what is possible in LIVEIA, a fledgling immersive technology that enables people to explore and develop new ways of understanding themselves and others. LIVEIA provides a means of visualizing and understanding other psychological phenomena in addition to those discussed here, such as wholeness versus fragmentation, integrity versus lack of integrity, closeness versus isolation, harmony versus lack of harmony, potentiality versus actualization, and the process by which experiences are assimilated and by which ideas are born and take shape. LIVEIA aims to effect understanding and positive transformation through imagery that works at an intuitive gut level by enabling people to visualize and play with their inner worlds.

LIVEIA grew out of earlier applications of optics to model cultural evolution (Gabora, 1999, 2002). More broadly, it is part of a research program aimed at developing a scientific framework for cultural evolution based on the hypothesis that what evolves through culture is the web of understandings that constitutes an individuals' ways of seeing and being in the world—sometimes referred to as 'worldviews' (Gabora 2000a, 2013; Gabora & Aerts, 2009). The goal behind LIVEIA was to complemente this objective, impersonal approach to the study of cultural evolution with an intuitive and engaging platform that facilitates the ability to visualize ones' worldview as part of an evolving tapestry of interacting worldviews, and prompt micro-moments of reflection on one's myriad ways of being and relating affect, however slightly, the process by which human culture evolves.

The research program opens further avenues of investigation. One such avenue is to using LIVEIA to better understand and track creative processes, and test theories about how creativity works (Gabora, 2000b). Another avenue involves getting a better grip on what is meant by the term inner light, for exemple by assessing the extent to which people agree in their assessments of the degree to which someone exudes (or obstructs) inner light. This could be accomplished using a modified version of a research protocol that has been used to assess the extent to which there is agreement amongst people's assessments of the degree of authenticity in creative performances (Henderson & Gabora, 2013).

The project is at an early stage. We are currently seeking collaborators in computer graphics, optics, and digital art to assist with building and testing an enhanced prototype.